# NODE-CONTEXT NETWORK CLUSTERING USING PARAFAC TENSOR DECOMPOSITION

Andri Mirzal and Masashi Furukawa
Graduate School of Information Science and Technology, Hokkaido University,
Kita 14 Nishi 9, Kita-Ku Sapporo 060-0814, Japan
Email: [andri, mack]@complex.eng.hokudai.ac.jp

## ABSTRACT

We describe a clustering method for labeled link network (semantic graph) that can be used to group important nodes (highly connected nodes) along with their relevant link's labels by using PARAFAC tensor decomposition. In this kind of network, the adjacency matrix can not be used to fully describe all information about the network structure. We have to expand the matrix into 3-way adjacency tensor, so that not only the information about to which nodes a node connects to but by which link's labels is also included. And by applying PARAFAC decomposition on this tensor, we get two lists, nodes and link's labels with scores attached to each node and labels, for each decomposition group. So clustering process to get the important nodes along with their relevant labels can be done simply by sorting the lists in decreasing order. To test the method, we construct labeled link network by using blog's dataset, where the blogs are the nodes and labeled links are the shared words among them. The similarity measures between the results and standard measures look promising, especially for two most important tasks, finding the most relevant words to blogs query and finding the most similar blogs to blogs query, about 0.87.

## 1  INTRODUCTION

The research on network clustering have a long tradition in computer science, especially on neighborhood-based network clustering, where the nodes being grouped together if they are in the vicinity and have a higher-than-average density of links connecting them [1-11]. Some examples of the real application of this network clustering are in parallel computing and distributed computation where the n number of tasks is divided into several processes that carried out by a separate program or thread running on one of m different processors [8].

In addition to the neighborhood-based network clustering, there is another clustering method that works on labeled link network, where the nodes are in the same group if they share the same or almost the same link's labels. In online auction networks this method can be used to find similar users [12], and by utilizing user's preferences in buying and selling activities, a recommendation system for effective and efficient advertisement can be proposed [13].

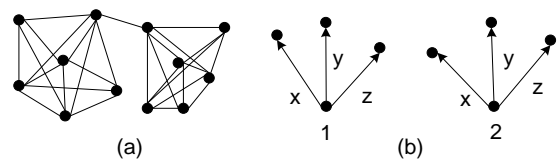

Figure 1. This figure shows the concept of clustering based on neighborhood (a), and based on link's labels (b). In (a), there are two clusters which well connected within cluster and only has one link between clusters. And in (b) node 1 and 2 are in the same group due to the similarity in their link's labels even though they are not connected at all.

These two clustering methods, however, can only group nodes. When the challenge to group the nodes with their most relevant link's labels comes, we have to utilize characteristic matrix , a nodes-versus-link's labels matrix, where the entries are the weight of the links. The further discussion about this can be found in [13].

But there are some situations where this approach is not suitable to be used, for example in the situation where we are only interested in finding the relevant link's labels for the most important nodes, nodes that have many links (in web network this will be web pages with many inlinks, but other networks like online auction network and international trading network, this can be nodes with many inlinks and/or outlinks [13]). This task is not trivial, for example in the web, the famous web pages more likely to attract many viewers, so that the ability to group the important pages with their relevant anchor text in the hyperlinks has real advantage [14].

To provide the natural way of grouping important nodes along with their relevant labeled links, we follow Kolda's works [14, 15]; first constructing adjacency tensor of the labeled link network, where the first and second axis are the nodes and the third axis is the link's labels (an example shown in fig. 2). And then apply tensor decomposition to get the node's authority and hub vectors along with link's labels' vectors for each *R* decomposition groups (see eq. (1) and fig. 3). And because the results produce ranking scores for both nodes and link's labels, we can sort these score vectors in descending order to get clustering of important nodes and their relevant labeled links for each R decomposition groups.

In this paper, instead of using real labeled link network, a problem that has already being discussed in [14, 15], we decide to use blogs dataset and form labeled link network by taking blogs as the nodes and shared words as the link's labels. This work has some possible real applications. For example because the similarity measures for task 1 (finding the most relevant blogs to the words query), and task 3 (finding the most similar blogs to the blogs query) give promising results (see table 4), one can build blogs search engine that provides users with ability to find the most relevant blogs to the words query and to find the most similar blogs to the blogs query. And because the results provide us with blog-word grouping, the engine can display not only the blogs but also the relevant words. Also one can also build visualization service that shows the blogs with its relevant shared words.

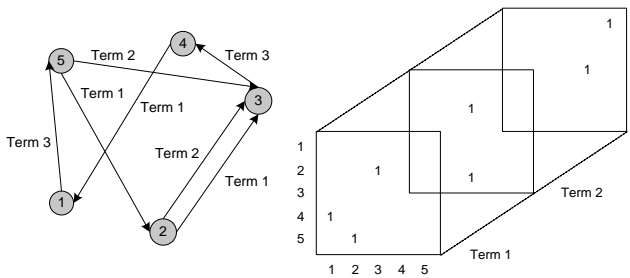

Figure 2. (a) The labeled link network with five nodes and three terms, and (b) its adjacency tensor

## 2 PARAFAC DECOMPOSITION

The PARAFAC tensor decomposition is higher-order analogue to the matrix singular value decomposition (SVD), but the singular vectors produced by PARAFAC are not generally orthonormal as the case in SVD [15].

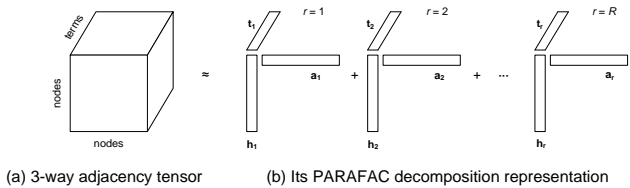

Figure 3. (a) Network's adjacency tensor, and (b) its R rank-1 PARAFAC decomposition

The PARAFAC decomposition approximates a tensor by the sum of $R$ rank-1 outer products of vectors $\mathbf{h}_r$, $\mathbf{a}_r$, and $\mathbf{t}_r$ as shown by fig. 3. Vector $\mathbf{h}_r$ is the corresponding hub vectors, $\mathbf{a}_r$ is the corresponding authority vectors, and $\mathbf{t}_r$ is the corresponding term vectors for each rank $r$.

PARAFAC decomposition of the adjacency tensor $\mathbf{X}$ can be written as (following Kolda et al [15]):

$$\mathbf{X} \approx \lambda \langle \mathbf{H}, \mathbf{A}, \mathbf{T} \rangle \equiv \sum_{r=1}^{R} \lambda_r \mathbf{h}_r \circ \mathbf{a}_r \circ \mathbf{t}_r \quad (1)$$

where $\mathbf{H}$, $\mathbf{A}$, and $\mathbf{T}$ is the hub, authority and term matrices of $R$ rank-1 tensor $\mathbf{X}$ decomposition, $\circ$ is outer vectors product, and $\lambda_r$ ($\lambda_1 \geq \lambda_2 \geq \cdots \geq \lambda_R$) is the weight for each group $r$. $\mathbf{H}$, $\mathbf{A}$ and $\mathbf{T}$ are formed by arranging vectors $\mathbf{h}_r$, $\mathbf{a}_r$, and $\mathbf{t}_r$ such that:

$$\mathbf{H} = [\mathbf{h}_1 \ \mathbf{h}_2 \ \cdots \ \mathbf{h}_r], \ \mathbf{A} = [\mathbf{a}_1 \ \mathbf{a}_2 \ \cdots \ \mathbf{a}_r], \ \text{and} \ \mathbf{T} = [\mathbf{t}_1 \ \mathbf{t}_2 \ \cdots \ \mathbf{t}_r] \quad (2)$$

To calculate PARAFAC decomposition, greedy PARAFAC algorithm is used [14].

## 3 DATA AND PREPROCESSING

We downloaded the blogs and its contents from technorati's[1] most popular and favorite blogs lists on November 27th, 2007. The number of blogs is 151 and the number of shared words is 704 (180 words after irrelevant words, like stop words, unrecognized words, and words that don't have clear meaning, are filtered out).

Data preprocessing manipulates the blogs and their contents into labeled link network. The blogs itself don't connect to each others by any hyperlinks. The links are the shared words and a blog connect to the others if they share the same words. And because the shared words usually don't appear once in the blogs, the labeled links are weighted with the total number of shared words' appearance in all blogs that share those words. Fig. 4 describes the process of constructing labeled links network from blogs dataset.

The blogs and shared words can be represented mathematically by characteristic matrix (see fig. 4 (b)). From this matrix, we can form bipartite graph (fig. 4 (c)), where the nodes are blogs and shared words. And then we can manipulate this bipartite graph into labeled link network (fig. 4 (d)), which is the form that we need to apply PARAFAC decomposition into its adjacency tensor (fig. 4 (e)). Because the network is constructed from bipartite graph, the result is undirected, so each frontal slices of the adjacency tensor (adjacency matrix for each shared words) is symmetric matrix.

Algorithm in fig. 5 is used to create adjacency tensor from characteristic matrix.

## 4 EXPERIMENT RESULTS

We decompose the adjacency tensor into 2, 4, …, 14 groups. Our code is written in MATLAB by using MATLAB Tensor Toolbox[2] [16] and run in a notebook with Mobile AMD Sempron processor 3000+ with 480 MB DDRAM. The maximum number of groups, 14, is not chosen but is the maximum number that our computer can process due to the memory limitation. The computational time increases rapidly as the number of groups increases, with approximately 15 minutes for the 14-group decomposition. Table 1 and 2 show the result of two-group and four-group decomposition.

---

[1] http://technorati.com/pop/blogs/
[2] http://www.models.kvl.dk/source/nwaytoolbox

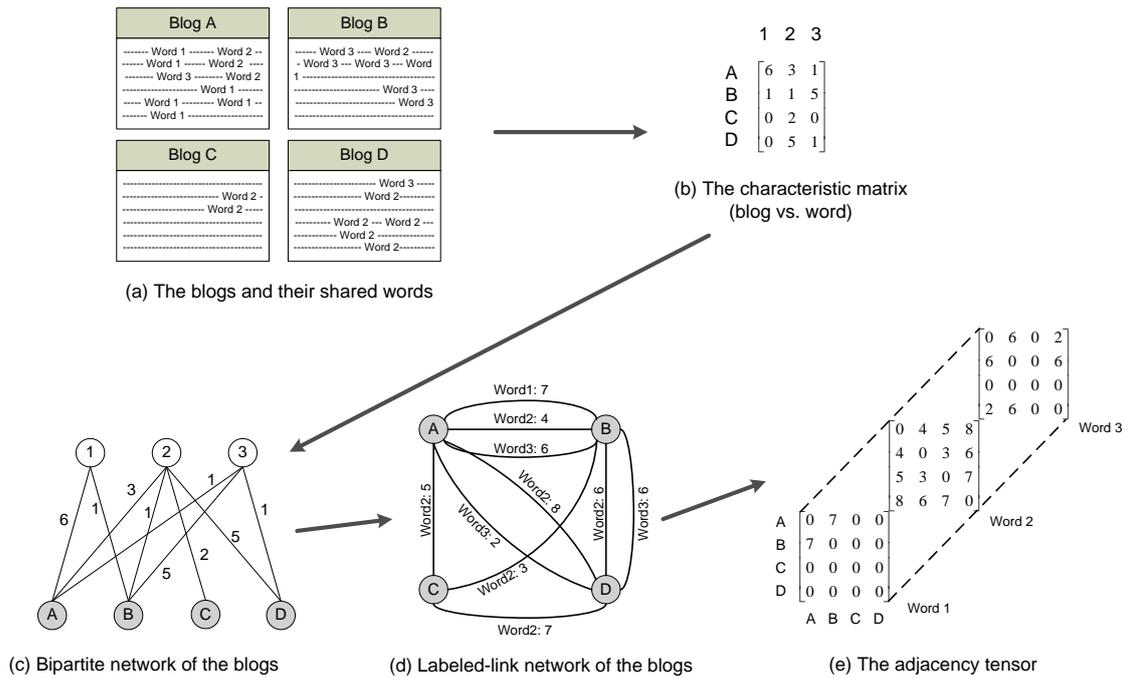

Figure 4. The labeled link network construction process from blogs dataset

| **In** | : Blog's Characteristic Matrix **C** | |
|---|---|---|
| **Out** | : Blog's Adjacency Tensor **X** | |
| $K$ = number of column of **C** | $\mathbf{Y} = \mathbf{X}$ |
| $I$ = number of row of **C** | for $k$ = 1, 2, ..., $K$, do |
| for $k$ = 1, 2, ..., $K$, do |   for $j$ = 1, 2, ..., $I$, do |
|   for $i$ = 1, 2, ..., $I$, do |     for $i$ = 1, 2, ..., $I$, do |
|     **X**($i$, :, $k$) = **C**(:, $k$); |       if $X(i, j, k)$ != 0 |
|     $X(i, i, k) = 0$; |         $X(i, j, k) = X(i, j, k) + Y(j, i, k)$; |
|     if $C(i, k) == 0$ |       end do |
|       **X**($i$, :, $k$) = 0; |     end do |
|   end do |   end do |
| end do | delete **U**, **Y** |

Figure 5. Algorithm to manipulate the characteristic matrix into adjacency tensor[3]

## 5 RESULTS ASSESSMENT

Two papers by Kolda et. al. [14, 15] that introduce the using of PARAFAC tensor decomposition to group web pages along with their anchor text, don't provide the assessment of the results, they only qualitatively state that tensor decomposition produces good results compare to the results by authoritative scores of HITS. We will move further here to assess the quality of tensor decomposition by using similarity measure, cosine criterion, between the results and standard measure results. And because decomposition produces two scores, blog and shared word scores, for each groups, we have to use query vectors that ask both decomposition results and standard measures for relevant/similar blogs or shared words. Because the query vectors and the groups to be found can be blogs or shared words, there are four possibilities in the query - result as described in table 3.

---
[3] Colon symbol (:) is an indexing notation that denotes full range of a given index

Table 1. Two-group decomposition

*First group*

| Blog | Score | Word | Score |
|---|---|---|---|
| A Consuming Experience | 2916.1 | Blog | 0.90727 |
| The Viral Garden | 1966.3 | those | 0.11279 |
| Search Engine Land | 1286.4 | Free | 0.11032 |
| Quartz Mountain Weblog | 1043.6 | video | 0.09926 |
| Tech Gadget Blog | 889.5 | world | 0.096373 |
| The thinking blog | 838.23 | Right | 0.095685 |
| i Thought, therefore i Blog | 826.66 | Top | 0.086629 |
| Bloggers Blog | 812.21 | news | 0.084289 |
| Neil Gaiman's Journal | 783.03 | media | 0.083858 |
| India PR Blog | 780.75 | online | 0.082614 |
| ... | ... | ... | ... |

*Second group*

| Search Engine Land | 2114.7 | google | 0.86815 |
|---|---|---|---|
| Google Operating System | 1668.4 | search | 0.44746 |
| A Consuming Experience | 1219.3 | engine | 0.063298 |
| Official Google Blog | 931.68 | social | 0.052491 |
| The Viral Garden | 631.51 | online | 0.047412 |
| Search Engine Roundtable | 599.15 | image | 0.042248 |
| Quartz Mountain Weblog | 577.57 | mobile | 0.041141 |
| Search Engine Watch Blog | 546.03 | news | 0.038686 |
| Blog Maverick | 544.02 | business | 0.03801 |
| Valleywag | 537.74 | information | 0.037634 |
| ... | ... | ... | ... |

Table 2. Four-group decomposition

*First group*

| Blog | Score | Word | Score |
|---|---|---|---|
| A Consuming Experience | 2228.3 | blog | 0.98113 |
| The Viral Garden | 1773.5 | marketing | 0.065005 |
| Quartz Mountain Weblog | 1124.9 | Free | 0.052199 |
| Search Engine Land | 1084.6 | Top | 0.048704 |
| Tech Gadget Blog | 961.13 | feature | 0.029986 |
| i Thought, therefore i Blog | 933.98 | online | 0.025942 |
| Bloggers Blog | 887.89 | media | 0.025532 |
| Neil Gaiman's Journal | 880.54 | reader | 0.024918 |
| The thinking blog | 878.92 | company | 0.023361 |
| India PR Blog | 873.41 | internet | 0.022821 |
| ... | ... | ... | ... |

*Second group*

| Search Engine Land | 2092.1 | google | 0.87334 |
|---|---|---|---|
| Google Operating System | 1686.8 | Search | 0.44775 |
| A Consuming Experience | 1036.7 | engine | 0.066004 |
| Official Google Blog | 937.78 | social | 0.049286 |
| Search Engine Roundtable | 647.05 | mobile | 0.040725 |
| Quartz Mountain Weblog | 581.14 | online | 0.035622 |
| Search Engine Watch Blog | 580.19 | yahoo | 0.03386 |
| The Viral Garden | 561.53 | image | 0.032817 |
| Valleywag | 552.72 | ads | 0.030336 |
| Blog Maverick | 542.83 | business | 0.029824 |
| ... | ... | ... | ... |

*Third group*

| Ask MetaFilter | 459.66 | blog | 0.97118 |
|---|---|---|---|
| TreeHugger Radio | 451.43 | google | 0.16145 |
| Techdirt | 450.12 | marketing | 0.058856 |
| Deadspin | 447.44 | search | 0.053256 |
| The Unofficial Apple Weblog | 446.67 | top | 0.048478 |
| Singapore Entrepreneurs | 446.59 | media | 0.048383 |
| Boing Boing | 443.47 | free | 0.046517 |
| 43 Folders | 442.48 | reader | 0.042519 |
| Topix.net Weblog | 441.5 | service | 0.03627 |
| Mashable! | 440.99 | social | 0.033971 |
| ... | ... | ... | ... |

*Fourth group*

| NewsBusters.org | 882.9 | world | 0.28811 |
|---|---|---|---|
| A Consuming Experience | 698.22 | right | 0.28704 |
| Search Engine Land | 668.38 | news | 0.28023 |
| The Corner | 658.24 | video | 0.24431 |
| Singapore Angle | 619.29 | america | 0.21115 |
| Gothamist | 579.5 | life | 0.1852 |
| we make money not art | 567.74 | media | 0.1562 |
| lifehack.org | 557.94 | free | 0.15519 |
| Blog Maverick | 551.35 | online | 0.14027 |
| the thinking blog | 542.45 | report | 0.13522 |
| ... | ... | ... | ... |

Table 3. Query - result relationship

| Query | Relevant/similar group to be found |
|---|---|
| Blogs | Blogs |
| Blogs | Shared words |
| Shared words | Blogs |
| Shared words | Shared words |

As the standard measure, because it has to be something that doesn't have or produces error and approximation values, we use blog's characteristic matrix **C** (an example is in fig. 4(b)). Before we move further to find relevant/similar group to the query, we have to calculate in advance the blog's similarity matrix **B** and shared word's similarity matrix **W**. Let $N$ be the number of blogs and $M$ be the number of shared words, matrix **B** is $N \times N$ matrix with its entries defined as:

$$B(i,j) = \cos \angle (\mathbf{C}(i,:), \mathbf{C}(j,:)), \quad 1 \leq i, j \leq N \quad (3)$$

and matrix W is $M \times M$ matrix with entries defined as:

$$W(p,q) = \cos \angle (\mathbf{C}(:,p), \mathbf{C}(:,q)), \quad 1 \leq p, q \leq M \quad (4)$$

The last thing to be considered before the similarity between tensor decomposition results and standard measure results being calculated is the query vectors. There are two query vectors, $N \times 1$ blog's query vector $\mathbf{q}_{blog}$, and $M \times 1$ word's query vector $\mathbf{q}_{word}$, where the entry is one if the blogs/words appear in queries and zero otherwise. Because we are only interested in knowing the quality of decomposition results in average, and not in evaluating specific cases, we set all entries of vector **q** to one.

## 5.1 Task 1: Finding the most relevant blogs to words query

For standard measure case, matrix **C** is being used.

$$\mathbf{b}_{standard} = \mathbf{C}\mathbf{q}_{word}, \quad \text{where} \quad \mathbf{q}_{word} = ones(M,1) \quad (5)$$

For tensor decomposition results we calculate:

$$\mathbf{m} = \mathbf{T}^T \mathbf{q}_{word} \text{ and } \mathbf{b}_{decomp.} = \mathbf{Hm} \quad (6)$$

The similarity between ranking vector of standard measure, $\mathbf{b}_{standard}$ and tensor decomposition results, $\mathbf{b}_{decomp.}$ is calculated using cosine criterion.

$$sim(\mathbf{b}_{standard}, \mathbf{b}_{decomp.}) = \cos \angle (\mathbf{b}_{standard}, \mathbf{b}_{decomp.}) \quad (7)$$

## 5.2 Task 2: Finding the most relevant words to blogs query

For standard measure case, matrix **C** is being used.

$$\mathbf{w}_{standard} = \mathbf{C}^T \mathbf{q}_{blog}, \quad \text{where} \quad \mathbf{q}_{blog} = ones(N,1) \quad (8)$$

For tensor decomposition results we calculate:

$$\mathbf{m} = \mathbf{H}^T \mathbf{q}_{blog} \text{ and } \mathbf{w}_{decomp.} = \mathbf{Tm} \quad (9)$$

The similarity between ranking vector of standard measure, $\mathbf{w}_{standard}$ and tensor decomposition results, $\mathbf{w}_{decomp.}$ is calculated using cosine criterion.

$$sim(\mathbf{w}_{standard}, \mathbf{w}_{decomp.}) = \cos \angle (\mathbf{w}_{standard}, \mathbf{w}_{decomp.}) \quad (10)$$

## 5.3 Task 3: Finding the most similar blogs to blogs query

For standard measure case, matrix **B** is being used.

$$\mathbf{b}_{standard} = \mathbf{B}\mathbf{q}_{blog}, \quad \text{where} \quad \mathbf{q}_{blog} = ones(N,1) \quad (11)$$

For tensor decomposition results we calculate:

$$\mathbf{m} = \mathbf{H}^T \mathbf{q}_{blog} \text{ and } \mathbf{b}_{decomp.} = \mathbf{Hm} \quad (12)$$

The similarity between ranking vector of standard measure, $\mathbf{b}_{standard}$ and tensor decomposition results, $\mathbf{b}_{decomp.}$ is calculated using cosine criterion.

$$sim(\mathbf{b}_{standard}, \mathbf{b}_{decomp.}) = \cos \angle (\mathbf{b}_{standard}, \mathbf{b}_{decomp.}) \quad (13)$$

## 5.4 Task 4: Finding the most similar words to words query

For standard measure case, matrix **W** is being used.

$$\mathbf{w}_{standard} = \mathbf{W}\mathbf{q}_{word}, \quad \text{where} \quad \mathbf{q}_{word} = ones(M,1) \quad (14)$$

For tensor decomposition results we calculate:

$$\mathbf{m} = \mathbf{T}^T \mathbf{q}_{word} \text{ and } \mathbf{w}_{decomp.} = \mathbf{Tm} \quad (15)$$

The similarity between ranking vector of standard measure, $\mathbf{w}_{standard}$ and tensor decomposition results, $\mathbf{w}_{decomp.}$ is calculated using cosine criterion.

$$sim(\mathbf{w}_{standard}, \mathbf{w}_{decomp.}) = \cos \angle (\mathbf{w}_{standard}, \mathbf{w}_{decomp.}) \quad (16)$$

## 5.5 The Similarity Measure Calculation Results

The total similarity measures for task 1 and 3 give good results, about 87%. And because in real situation users usually want to find set of blogs that match with their keywords (task 1, find the most relevant blogs to words query), and also the most similar set of blogs to the their favorite blogs (task 3, find the most similar blogs to blogs query) the high similarity measure for task 1 and 3 seems to be promising.

Also the relatively good result in task 3, finding the most relevant words to the blogs query means that tensor decomposition can also be used to give description of the contents of queried blogs.

The worst case is the task 4, about 63%, finding the most similar words to words query. But because in our opinion this task has no clear purpose, we can exclude it from our discussion.

From all groups that adjacency tensor being decomposed into, the 4-group gives the best average similarity measure. But unfortunately we still don't know how to predict this in advance.

Table 4. The summary of similarity measure

|        | Group 2 | Group 4 | Group 6 | Group 8 | Group 10 | Group 12 | Group 14 | Av.    |
|--------|---------|---------|---------|---------|----------|----------|----------|--------|
| Task 1 | 0.8854  | 0.8748  | 0.8772  | 0.8477  | 0.9077   | 0.8561   | 0.8839   | **0.8761** |
| Task 2 | 0.8715  | 0.7924  | 0.8244  | 0.6905  | 0.9223   | 0.7325   | 0.7994   | 0.8047 |
| Task 3 | 0.8421  | 0.9461  | 0.8384  | 0.9460  | 0.7329   | 0.9426   | 0.8812   | **0.8756** |
| Task 4 | 0.4255  | 0.7666  | 0.7257  | 0.7155  | 0.6320   | 0.5674   | 0.5660   | 0.6284 |
| Av.    | 0.7561  | **0.8450** | 0.8164 | 0.7999 | 0.7987  | 0.7747   | 0.7826   | **0.7962** |

## 6  DISCUSSION

Actually there are some questions remain here. For example, how to choose the best number of groups for any adjacency tensors, how to associate the blogs that always placed in the bottom part with their shared words, etc. These questions, even though are very interesting, can not be answered right now. And the mechanism for filtering out the irrelevant items and the system that has ability to recognize phrases in addition to the words have not applied to this work. We will address these all remaining tasks in the future research.